\documentclass[a4paper,12pt]{article}

\usepackage[pdftex]{graphicx}
\usepackage{authblk}
\usepackage{amssymb}
\usepackage[nolist]{acronym}
\usepackage{comment}

\title{Machine Learning in Gamma Astronomy}

\author[1]{A.P.Kryukov\thanks{kryukov@theory.sinp.msu.ru}}
\author[1]{A.P.Demichev\thanks{demichev@theory.sinp.msu.ru}}
\author[2]{V.A.Ilyin\thanks{ilyin0048@mail.ru}}
\affil[1]{\small{\textit{Lomonosov Moscow State University, Skobeltsyn Institute of Nuclear Physics,}}\\ \small{\textit{1(2), Leninskie gory, GSP-1, Moscow, 119991, Russia}}}
\affil[2]{\small{\textit{National Research Centre "Kurchatov Institute", 123182 Moscow, Russia}}}

\date{}

\begin{document}

\maketitle

\begin{abstract}
The purpose of this paper is to review the most popular deep learning methods used to analyze astroparticle data obtained with Imaging Atmospheric Cherenkov Telescopes and provide references to the original papers.
\end{abstract}

\section{Introduction}\label{intro}

\label{sec:Int}

Imaging Atmospheric Cherenkov Telescopes (IACT) register extensive air showers (EAS) generated by gamma rays and cosmic rays (charged particles) when they interact with the atmosphere. These events are images recorded by IACT's highly sensitive camera, which consists of hundreds of photomultiplier tubes. Therefore, the main task of analyzing IACT data is the ability to distinguish between these two types of EAS. In addition, other  properties of the primary high-energy particle, such as energy and direction of arrival, may be determined too.

In recent years, deep learning methods have made significant progress in identifying gamma events and reconstructing their features. 
The main purpose of this paper is to provide an overview of works that use deep learning to analyze IACT data.
Note that the principles of deep learning methods themselves are not considered here. 
They can be found, for example, in the books \cite{aggarwal2018neural, goodfellow2016deep} and in the reviews \cite{shrestha2019review, hatcher2018survey, dong2021survey}. 
The review also does not cover machine learning methods that do not use deep learning, such as random forests \cite{albert2008implementation} or boosted decision trees \cite{ohm2009gamma, becherini2011new}; see also \cite{sharma2014gamma} and references therein. 
As a rule, we do not provide details of the physical results obtained in the papers under review. 
They are closely related to the physics of EASs, and their discussion would overload the text. 
We assume that the reader interested in the methods under consideration will find further details in the original papers.

There are several reviews devoted to this discussed topic. 
The review \cite{dAmico2022statistical} aims to summarize the most common statistical tools for IACT data analysis. 
The short review \cite{fanelli2020machine} focuses on advances in detector design and calibration using machine learning methods. 
The recent review \cite{malyshev2023analysis} is mainly devoted to traditional methods of analysis in gamma-ray astronomy and very briefly mentions deep learning methods.

This paper is a shortened version of the extensive review \cite{demichev2024using}. 
It also includes a number of new papers that were published after the review was written. This will allow the reader to better understand the state of the art in this field.

\section{Particle type classification}
\label{ssec:PTC}

\begin{figure*}
	\centering
	\includegraphics[scale=0.3]{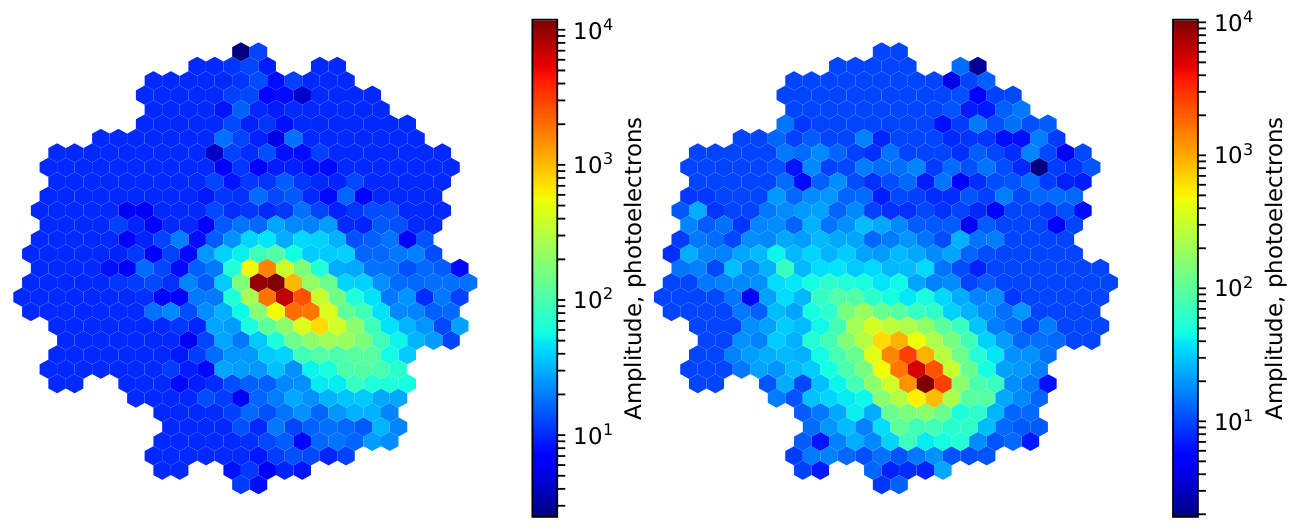}
	\caption{\label{fig:EAS_image} Examples of simulated EAS images in an IACT camera for the TAIGA experiment (on the left: for an EAS initiated by a gamma ray, on the right: for an EAS from a charged particle (proton)).}
\end{figure*}

A difficulty of ground-based gamma astronomy is that gamma-ray showers account for a small fraction of the total number of observed EASs.
The background from showers induced by charged particles exceeds signal events by a factor of $10^4$. The ability to distinguish between gamma-ray and cosmic-ray events is one of the main factors determining the feasibility of studying gamma-ray sources using IACT. Gamma-ray showers are caused by electromagnetic processes, while in cosmic-ray showers hadronic processes play an important role. This difference is reflected in the structure of the shower image in the IACT camera and can be used to classify events (see Figure~\ref{fig:EAS_image}). Gamma-ray event images are more elongated and directed toward the center of the camera, while proton events are less elongated and have random directions. 
One of the first papers on the use of deep learning in gamma-ray astronomy is the paper \cite{reynolds1993neural}, which used a network with only one hidden layer.
It was shown that neural networks can separate gamma and proton events.

The paper \cite{nietoCastano2017exploring} presents the results of a study on the applicability of deep learning as a method for classifying \ac{eas} events within the \ac{cta} project \cite{acharya2013introducing}.
For this purpose, \ac{mc} event simulations in \ac{cta} were used.
The \acp{eas} simulations were performed using the CORSIKA package \cite{heck1998corsika}, and the simulation of telescopes and cameras were performed using specialized software for \ac{cta}.
In~\cite{nietoCastano2017exploring} two well-known neural network architectures were used: ResNet50 \cite{he2016deep} and Inception V3 \cite{szegedy2016rethinking} (both are available as applications in Keras). 
Thanks to the use of batch normalization, optimization of layer sizes, and careful balancing of the width and depth of the network, Inception V3 shows good performance. 
ResNet50 uses a residual mapping approach, which is implemented as shortcut connections through convolutional layers. 
This allows very deep networks to be trained more efficiently.
The Inception V3 neural network showed slightly better, although still close to ResNet50, accuracy rates. 
The paper demonstrated that deep learning methods can be used to classify \ac{iact} images without any prior parameterization and any assumptions about the physical processes in EAS. 
The classification accuracy in the tested models depends on the energy of the primary particles of \ac{eas}. For the Inception V3 architecture, the accuracy increases from 81.4\% to 91.6\% in the energy range from 0.1 to 100 TeV.

The paper \cite{shilon2019application} presents the results of the background (hadron-induced showers) rejection  and reconstruction of the \ac{eas} direction of arrival by analyzing \ac{iact} data with the help of combination of \acp{cnn} applied to images from four telescopes of the H.E.S.S. experiment \cite{ashton2020nectar}. An interesting feature of this paper is the first attempt to use recurrent neural networks (RNNs) in combination with \ac{cnn} to analyze image sequences time-ordered by triggers of each of the four \ac{iact} telescopes. 
This approach helped to solve some technical problems. However, the authors of \cite{shilon2019application} did not find sufficiently convincing arguments for its further use. In particular, the computational requirements in this case are significantly higher.

The problem of converting \ac{iact} data using a hexagonal packing of photomultiplier tubes in cameras to the rectangular grid format of a convolutional neural network (CNN) is the subject of \cite{holch2017probing, nieto2019studying}. These papers present and compare several strategies for solving this problem. 
Their conclusion is that the values of the main quality criteria of neural networks, such as accuracy and ROC/AUC score, coincide within errors for all the methods studied. 
Somewhat worse results were obtained only for the nearest interpolation. 
One more method is based on tranformation to the regular square grid by using oblique coordinates with the angle $60^\circ$ (see, e.g. \cite{postnikov2019gamma}). 
A very interesting approach to solving the hexagonal lattice problem is to apply special convolution operations to such pixel geometry. To do this, one can use the IndexedConv package \cite{jacquemont2019indexed}, based on the PyTorch framework \cite{ketkar2017deep}. 
This package allows the user to apply convolution to any pixel organization defined by a matrix whose elements are non-zero only for adjacent pixels.

In \cite{parsons2020background}, high-performance hadronic background rejection algorithm was developed based on convolutional recurrent neural networks (CRNNs) using data from several telescopes (stereoscopic mode). It permits significantly increase the efficiency of background rejection in the energy range from 0.1 to 100~TeV.  The paper demonstrated that CRNNs improve the hadronic background rejection of about 20–25\% compared to using the conventional algorithms based on Hillas parameters. Note that when only simulated data are used, the quality of the background rejection is even higher (20\% for RNN with Hillas parameters as the input and 60\% for CRNN with \ac{eas} images as the input). The problem of background suppression caused by primary electrons was also studied in the paper~\cite{spencer2021deep}. It is also emphasized that the use of \acp{cnn} can provide a direct event classification method that takes into account all the information contained in the EAS images. This may allow online data processing. 
In the paper \cite{feng2016analysis} deep learning methods apply not only to the analysis of gamma rays, but also to other particles, such as muons.

In \cite{riquelme2023deep}, a self-supervised learning approach is explored. It allows using of new unlabeled images  of real data to improve the performance of the classifier. This technique is based on the so-called pseudo-labeling strategy, which enables training on labeled and unlabeled data. 

A distinctive feature of \cite{de2022deep} is that it not only poses the problem of classifying already known primary particles like gammas and protons that generate \ac{eas}, but also the search for unknown particles. The authors propose to use autoencoders based on convolutional layers (CAE). The main idea is that the CAE is trained on simulated \ac{mc} events initiated by known particles. Then, when presented with a real image \ac{iact}, such an autoencoder reconstructs it well in the case of a known primary particle and distorts it in the case of an unknown particle. This may be a signal of the presence of anomalous events in the cosmic ray flux.

\begin{figure}
	\centering
	\includegraphics[scale=0.5]{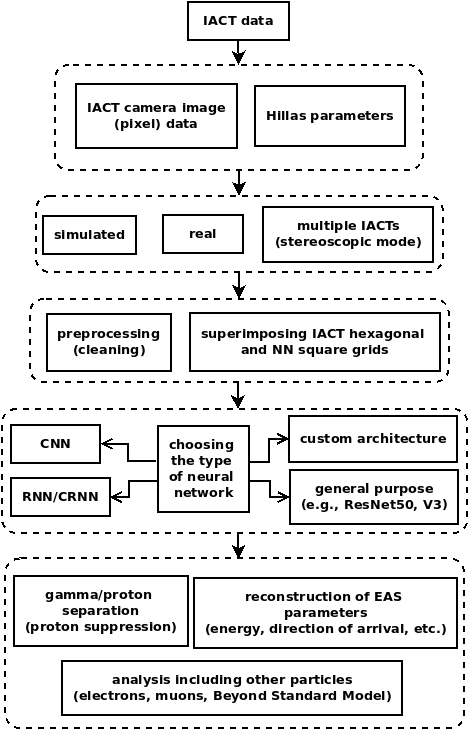}
	\caption{\label{fig:Diagram1} The key components of the deep learning based IACT data processing.}
\end{figure}

The key components of the deep learning based IACT data processing for particle type classification are presented in Figure~\ref{fig:Diagram1}. 

\section{Reconstruction of EAS parameters}
\label{ssec:DEA}

The aim of \cite{mangano2018extracting} is to test the possibility of using \acp{cnn} to extract gamma-ray events from the background and reconstruct the EAS parameters for the \ac{cta} setup.
In contrast to the previously considered papers, here for the first time \ac{cnn} was used for processing \ac{cta} simulated data not only to classify particles (rejection of the hadron background), but also to reconstruct the gamma-ray parameters. To reduce the complexity of processing, the main attention in the paper was paid to the processing of showers recorded by four telescopes simultaneously (stereoscopic mode). 
Using more than one telescope greatly improves the ability to reconstruct the properties of the primary particles, since the same \ac{eas} is registered from several locations.
Also, to simplify the analysis, four images were combined into one image by summing the pixel values. 
The transformation from the hexagonal grid of the \ac{cta} cameras to the rectangular one was carried out by the oversampling method. 
The authors found the results of \acp{cnn} to be promising, although they are still not as good as the results of existing algorithms based on Hillas parameters. 
The main advantages of \acp{cnn}s over conventional algorithms are that they do not require additional physical assumptions and require minimal data preprocessing. 
A similar approach was used in \cite{postnikov2019gamma, polyakov:2021use} for data simulated for the telescopes of the TAIGA project \cite{budnev2020taiga}. 

In \cite{postnikov2019gamma}, \acp{cnn} based on two open source machine learning libraries, PyTorch \cite{ketkar2017deep} and TensorFlow \cite{abadi2016tensorflow}, were used to investigate the possibility of rejecting hadron background events. The networks built on the basis of both platforms showed approximately the same results, which do not exceed the quality obtained by means of the Hillas parameters and machine learning methods without neural networks. 

The general conclusion is that the use of CNN-based methods is promising, but requires further improvements. 
More complex neural networks such as ResNet, GoogLeNet and DenseNet, combined with the joint processing of images from multiple telescopes, lead to more promising results~\cite{polyakov:2021use, gres2022energy}.

\begin{figure}
	\centering
	\includegraphics[scale=0.5]{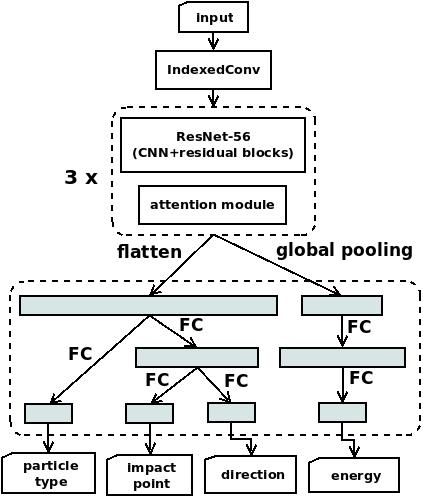}
	\caption{\label{fig:Diagram2} The simplified $\gamma$-PhysNet architecture.}
\end{figure}

The aim of papers~\cite{jacquemont2020single, jacquemont2021multi} is to separate gamma rays from charge cosmic rays and restore their parameters (energy and direction of arrival) using multi-task deep learning methods. 
The neural network is applied in the context of one LST-1 telescope, the first prototype of the \ac{cta} telescope built in the Northern Hemisphere (La Palma, Canary Islands). 
The $\gamma$-PhysNet system presented in the paper is a multi-task neural network model that performs full event reconstruction with a single deep neural network, see Figure~\ref{fig:Diagram2}. 
It consists of two parts. 
The first one is a very deep \ac{cnn} (ResNet 56), which plays the role of an encoder and is supplemented with blocks with an attention mechanism (see, for example, \cite{aggarwal2018neural} and references therein as well as the famous work \cite{vaswani2017attention}). 
The attention units help the model focus on the relevant part of the feature map \cite{sun2020saunet} to extract important features of the input data. 
This representation is then passed to the second part of the network, which is a multi-task unit. 
This unit is designed to reconstruct the energy, extract gamma rays from the background, and reconstruct the direction of the EAS axis and the arrival point.
Multi-task learning helps to improve the generalization properties of the model and, therefore, the quality of parameter reconstruction.
The $\gamma$-PhysNet system implements a  method of upload data~\cite{jacquemont2019indexed} that allows deep learning to be applied directly to the LST-1 images with hexagonal grid.
The performance of the system was evaluated using the results of the \ac{mc} simulations for LST-1.
The performance of the $\gamma$-PhysNet system was compared with the conventional method based on the Hillas parameters and classification by the Random Forest method \cite{albert2008implementation}. 
The angular and energy resolution show that $\gamma$-PhysNet outperforms this conventional method. 
The improvement is noticeable at the lowest energies (below 100~GeV). 
In particular, $\gamma$-PhysNet improves the accuracy of reconstructing the EAS axis direction to $0.3^\circ$.

Let us focused in more detail on two main features that distinguish the $\gamma$-PhysNet system \cite{jacquemont2020single, jacquemont2021multi} from other approaches to the analysis of \ac{iact} data, namely on multi-task learning and attention mechanism. 

The main idea of multi-task learning (MTL) \cite{zhang2021survey} is to improve the generalization capabilities of a neural network by exchanging representations between related tasks.
In other words, MTL helps the neural model focus on features that are relevant to all tasks. 
In MTL, the model is trained simultaneously using a partially shared representation for all tasks. 

In the most cases MTL models use hard sharing architecture. 
In this case, the first layers of the network are used simultaneously by all tasks. Then the network branches out and each branch is used for its own task.
In soft-sharing architectures, separate networks are trained for each task, but they share some information with each other.

Attention is a mechanism that helps the deep learning model to focus on essential features based on a specific context with the help of trainable weights. 
It originated in the field of natural language processing, but was generalized in \cite{parmar2018image} for image generation and later for image classification and object detection. 

There are two models of attention: soft and hard.
In the first case, the network accesses all the data it has access to, but the importance of this data is different.
This makes a neural network with this attention property more accurate, but not faster than regular RNNs.

In the second case, the network accesses only a part of all existing data, which improve of the accuracy and speed of calculations. 
However, such a network is significantly more complex in terms of its training, since it ceases to be continuous, and therefore differentiable. 
The paper~\cite{jacquemont2021multi} explore three variants of networks with attention: ``Self-attention'', ``Squeeze-and-Excitation'' and ``Dual Attention''. 
As already noted, the comparison of $\gamma$-PhysNet with the Hillas and Random Forest parameter-based method shows that neural networks using MTL and attention mechanism have better  performance and accuracy of \ac{iact} data analysis. 
It is possible to obtain more detailed energy spectra of gamma sources. 
Improvements in angular resolution and classification quality increase the signal-to-noise ratio, allowing detection of significantly weaker sources. 
In addition, it is important for the comparative study of point and extended sources. The attention mechanism was also successfully used for \ac{iact} data analysis in the papers~\cite{miener2021reconstruction, miener2022performance, grespan2022deep}.

The papers \cite{jacquemont2021first, abe2021analysis} are a continuation of the papers \cite{jacquemont2020single, jacquemont2021multi} and aim to assess how the results obtained on simulated data transfer to the analysis of real data.
In particular, the systematic learning error due to the difference between simulated and real data is discussed. 
Of particular importance is the difference between these two data types related to night sky background. 
This problem was solved by adding noise to the simulated data used to train the model. 
The Hillas+Random Forest method trained on simulated data is used as a reference method for both simulated and real data. 
The $\gamma$-PhysNet system, based on multi-task learning and an attention mechanism, was able to detect a signal from the Crab Nebula with a statistical significance of $14.3\,\sigma$, which outperforms the conventional approaches. 
In the paper~\cite{juryvsek2021full} the InceptionV3 model is used, the results of which were compared with boosted decision trees method. 
Overall, the networks performed better. 
The authors noted that decrease the quality of the \ac{mc} data leads to a noticeable decrease in the accuracy of the reconstruction of the parameters of the primary particles.

The authors of \cite{bylund2021studies} studied the possibility of using \acp{cnn} for the ALTO detector of the CoMET project\footnote{https://alto-gamma-ray-observatory.org}, which significantly simplified the procedure for reconstructing the primary properties of gamma rays, such as energy and maximum shower depth ($X_{max}$).
The main conclusion of \cite{bylund2021studies} is that the quality of \ac{eas} energy reconstruction by ordinary (that is, without RNN, LSTM and attention blocks) \acp{cnn} does not significantly exceed conventional approaches. 
This is consistent with the findings of other authors presented in this review. 
On the other hand, even \acp{cnn} \cite{bylund2021studies} performed better than conventional approaches in determining $X_{max}$. This means that \acp{cnn} are able to more accurately account for correlations between observed image features \ac{eas} and physical parameters.

The main issue studied in \cite{brill2019investigating} is the possible dependence of the performance of RNN networks (more precisely, CRNNs) on the method of ordering images of the same \ac{eas} from different telescopes (i.e., in the stereoscopic regime). 
To understand the impact of ordering on neural network performance, two CRNNs were trained to classify simulated gamma and proton images. 
The only difference in training was the change in the sequence of the input images. 
The control variant was a neural network trained on images ordered by telescope identification number. 
Another network was trained on images ordered by the \textit{Size} parameter (the overall intensity of the image). 
The quality of the two models in \cite{brill2019investigating} was about the same. The reference model was even slightly better. 
It achieved 80.6\% accuracy and 0.899 AUC, while the model with images ordered by \textit{Size} parameter achieved 80.2\% and 0.894, respectively.
Therefore, it was not possible to obtain confirmation that image sorting by the \textit{Size} parameter improves the efficiency of gamma-proton classification using the CRNN model. 
This leaves open the possibility that a different telescope image ordering could lead to improved CRNN performance. 
Although the \textit{Size} parameter is related to the \ac{eas} parameters, training on \textit{Size} ordered images has little effect on the quality of the model.

\begin{figure}
	\centering
	\includegraphics[scale=0.5]{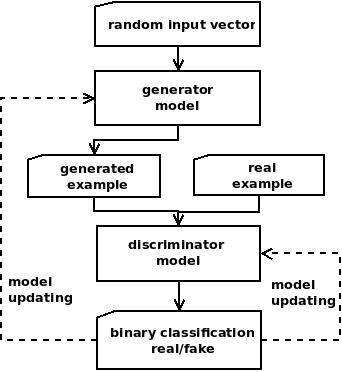}
	\caption{\label{fig:Diagram3} The general GAN architecture.}
\end{figure}

In \cite{dubenskaya:2021fast, dubenskaya:2021modeling, dubenskaya2021architecture} it was proposed to use generative adversarial networks (GAN) \cite{goodfellow2014generative} (Figure~\ref{fig:Diagram3}) as a replacement for the MC generator for fast modeling of gamma event images in \ac{iact}. 
The method is supposed to be used for data analysis within the TAIGA experiment \cite{budnev2020taiga}. 
Currently, modeling of event images for the \ac{iact} TAIGA project, as in almost all other similar projects, is carried out using the CORSIKA software based on the \ac{mc} methods. Additionally, the specialized program OPTICA-TAIGA \cite{postnikov2017hybrid} is used to simulate the response \ac{iact} and perform a full tracing of Cherenkov photons through the telescope optics.

The difficulty lies in the fact that the computational models underlying the physical processes are very resource intensive and require a lot of computational time. 
For some analysis purposes, such as reducing class imbalance using synthetic resampling \cite{johnson2019survey}, full model information is redundant. 
Therefore, less complex and more efficient generation methods can be used. 
The papers \cite{dubenskaya:2021fast, dubenskaya:2021modeling, dubenskaya2021architecture} show the possibility of using GAN for fast simulation of gamma/proton event images for the TAIGA-IACT experiment. 
It has been shown that using GANs allows one to quickly generate sets of new images whose parameter distributions are statistically indistinguishable from those of the images in the training set. 
The training was done using a sample of images obtained with the \ac{mc} simulation program, containing 25,000 gamma and 25,000 proton events.
Of particular interest are conditional GANs (cGANs), which allow images to be divided into several classes depending on the value of some image property, and then specify the desired class when generating new images.
In \cite{dubenskaya2022using}, the cGAN method was used to create images similar to those obtained in the TAIGA-IACT experiment with a predetermined spectrum in terms of the \textit{Size} parameter while in \cite{dubenskaya2024image} with a predetermined spectrum in terms of the energy.  
In \cite{polyakov2022using}, for the same purpose of generating images in an \ac{iact} camera with a given spectrum, a conditional variational autoencoder (cVAE) was used.
It was shown that GAN, cGAN and cVAE simulate proton and gamma events for the TAIGA-IACT experiment with a high degree of accuracy and reliability. 
It was shown that GAN, cGAN, and cVAE simulate proton and gamma events for the TAIGA-IACT experiment with a high accuracy and reliability. Most of the generated events are indistinguishable from those generated using the \ac{mc} method. 
The event generation rate using generative models is orders of magnitude higher than for the \ac{mc} method.

\section{Dedicated software for analyzing IACT data}
\label{sec:SSA}

The CTLearn software \footnote{The code is freely available on GitHub: https://github.com/ctlearn-project/ctlearn.} \cite{nietoCastano2019ctlearn} provides a backend for training neural networks for event reconstruction \ac{iact} using TensorFlow.
Input data can be loaded in three modes: mono (single images from a particular type of telescope); stereo (events recorded by multiple telescopes of the same type); and multi-stereo (events involving multiple types of telescopes).
The core module of the CTLearn software is the \textit{run\_model} module, which parses the configuration parameters, loads the data, and initializes the model.
The CTLearn framework also includes a number of helper scripts that provide a convenient way to present the results and plot the corresponding graphs.

\begin{figure}[bt]
	\centering
	\includegraphics[scale=0.45]{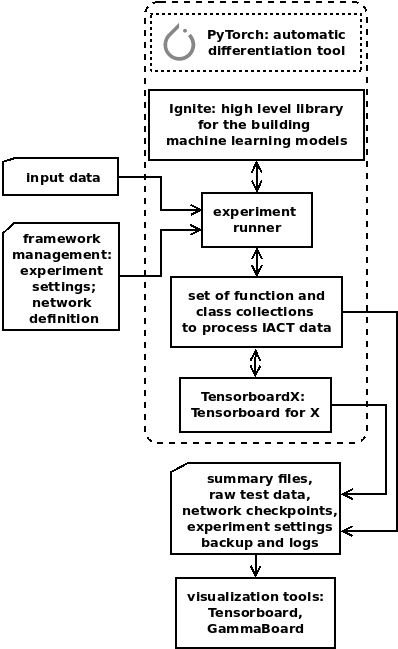}
	\caption{\label{fig:Diagram4} The simplified general architecture of GammaLearn.}
\end{figure}

The goal of the GammaLearn project \footnote{The code is freely available on Gitlab IN2P3: https://gitlab.in2p3.fr/gammalearn/gammalearn.} \cite{jacquemont2019gammalearn, vuillaume2019gammalearn} is to optimize neural networks by hyperparameters for the problems of gamma-ray and cosmic-ray classification and gamma-ray parameter reconstruction.
The software is based on the PyTorch framework and has an advanced set of tools that provide all the necessary functions for loading data sets, pre-processing data (filtering, augmentation, transformation), network training, validation and testing, monitoring the training process, and visualizing results.  
For example, GammaLearn contains the IndexedConv package, which provides convolution and pooling operations for input data (images) on any grid including a hexagonal grid.
The GammaBoard package provides a dashboard that displays various event reconstruction metrics \ac{iact}.
In addition, the GammaLearn environment provides useful tools for visualizing monitoring data (such as the weight distribution of a neural network during training or the GPU memory used), training metrics (such as the evolution of the error or accuracy function values).
A simplified general architecture of GammaLearn is depicted in Figure~\ref{fig:Diagram4}.

\section{Discussion and conclusion} 
\label{sec:Con}

The review of the papers on the application of deep learning methods for analyzing IACT data shows that their use is promising for both existing installations and future generation telescopes. 
An important advantage of neural networks is that they provide the ability to reconstruct the parameters of events associated with extensive air showers directly from images in IACT cameras, without requiring complex preprocessing steps. Convolutional neural networks (CNNs) which are the basis of the most popular of deep learning methods applied to image processing, demonstrate the high quality of IACT data processing. 
Future generation setups will consist of several or even many synchronized IACTs (stereoscopic mode). Thus, an approach based on combining CNNs with recurrent networks (RNNs) and LSTM cells allows processing series of images from different telescopes. 
Also note that, given that the EAS image (number of pixels fired) in an IACT camera usually occupies a relatively small part of the entire camera area, the use of neural networks with an attention mechanism can be productive.
Importantly, despite the difficulties of the training and optimization process, neural networks show good performance during execution, which makes them potentially suitable for real-time data analysis.

The widespread use of deep learning methods for analyzing IACT data will also be facilitated by the development of specialized computer platforms that provide users with convenient tools and automate the processes of creating and configuring the corresponding neural networks.

\subsection*{Acknowledgments}
The authors would like to thank the TAIGA collaboration for support and data provision. The work was carried out using equipment provided by the MSU Development Program.

\subsection*{Funding}
This study was supported by the Russian Science Foundation, grant no. 24-11-00136. 

\begin{acronym}[ICANN]
	\acro  {iact} [IACT]   {Imaging Atmospheric Cherenkov Telescope}
	\acro  {eas}  [EAS] {extensive air shower}
	\acro  {cta}  [CTA]   {Cherenkov Telescope Array}
	\acro  {bdt}  [BDT]  {boosted decision trees}    
	\acro  {nsb}  [NSB]  {night sky background}
	\acro  {msc}  [MSCW/MSCL]  {mean scaled width/length}   
	\acro  {mc}   [MC]  {Monte Carlo}
	\acro  {cnn}  [CNN]  {convolutional neural network}  
	\acro  {doa}  [DOA]  {direction of arrival}	
\end{acronym}

\end{document}